\documentclass[nofootinbib,twocolumn,superscriptaddress,longbibliography]{revtex4-1}

\usepackage{enumitem}
\usepackage{bm}
\usepackage[normalem]{ulem}
\usepackage{color}
\usepackage{enumerate}
\usepackage{amsfonts}
\usepackage{amssymb}
\usepackage{amsmath}
\usepackage{gensymb}
\usepackage{latexsym}
\usepackage{graphics}
\usepackage{graphicx}
\usepackage{epsfig}
\usepackage{multirow}
\usepackage{comment}
\usepackage{xcolor}
\usepackage{afterpage}
\usepackage{makecell} 
\usepackage{booktabs}
\usepackage{blindtext}
\usepackage{tabularx}
\usepackage[colorlinks=true,urlcolor=blue,anchorcolor=blue,citecolor=blue,filecolor=blue,linkcolor=blue,menucolor=blue,linktocpage=true,pdfproducer=medialab,pdfa=true]{hyperref}
\usepackage[utf8]{inputenc}
\usepackage{url}

\newcommand{\eV}{\textrm{eV}}

\renewcommand{\vec}[1]{\textbf{#1}} 
\renewcommand{\hat}[1]{\textbf{#1}}

\newcommand{\vobsvec}{\vec{v}_{\text{obs}}}
\newcommand{\vsun}{v_{\odot}}
\newcommand{\vsunvec}{\vec{v}_{\odot}}

\newcommand{\vearthvec}{\vec{v}_{\oplus}}

\newcommand{\vlabvec}{\vec{v}_{\rm lab}}
\newcommand{\sigmav}{\sigma_{\rm v}}

\newcommand{\grada}{\boldsymbol{\nabla a}}
\newcommand{\rhodm}{\rho_{a}}
\newcommand{\omegaE}{\omega_{\oplus}}
\newcommand{\axis}{m}
\newcommand{\ma}{m_a}

\newcommand{\ts}{\mathrm{TS}_{\rm max}}
\newcommand{\texp}{T_{\rm exp}}

\newcommand{\blueish}{purple}
\newcommand{\reddish}{orange}
\newcommand{\greenish}{green}

\newcommand{\dd}{\mathrm{d}}

\newcommand{\be}{\begin{eqnarray}}
\newcommand{\ee}{\end{eqnarray}}
\newcommand{\bmat}{\left(\begin{matrix}}
\newcommand{\emat}{\end{matrix}\right)}
\newcommand{\el}{\nonumber\\}

\newcommand{\Eq}[1]{Eq.~(\ref{#1})}
\newcommand{\Eqs}[2]{Eqs.~(\ref{#1}) and (\ref{#2})}
\newcommand{\Sec}[1]{Sec.~\ref{#1}}

\newcommand{\Fig}[1]{Fig.~\ref{#1}}

\let\Ref\undefined 
\newcommand{\Ref}[1]{Ref.~\cite{#1}}

\begin{document}

\title{Stochastic Properties of Ultralight Scalar Field Gradients}

\author{Mariangela Lisanti}
\affiliation{Department of Physics, Princeton University, Princeton, NJ 08544}

\author{Matthew Moschella}
\affiliation{Department of Physics, Princeton University, Princeton, NJ 08544}

\author{William Terrano}
\affiliation{Department of Physics, Princeton University, Princeton, NJ 08544}

\date{\today}

\begin{abstract}
  Ultralight axion-like particles are well-motivated dark matter candidates that are the target of numerous direct detection efforts.
  In the vicinity of the Solar System, such particles can be treated as oscillating scalar fields.
  The velocity dispersion of the Milky Way determines a coherence time of about $10^6$ oscillations, beyond which the amplitude of the axion field fluctuates stochastically.
  Any analysis of data from an axion direct detection experiment must carefully account for this stochastic behavior to properly interpret the results.
  This is especially true for experiments sensitive to the gradient of the axion field that are unable to collect data for many coherence times.
 Indeed, the direction, in addition to the amplitude, of the axion field gradient fluctuates stochastically.
  We present the first complete stochastic treatment for the gradient of the axion field, including multiple computationally efficient methods for performing likelihood-based data analysis, which can be applied to any axion signal, regardless of coherence time.
  Additionally, we demonstrate that ignoring the stochastic behavior of the gradient of the axion field can potentially result in failure to discover a true axion signal.
\end{abstract}

\maketitle

\section{Introduction}

If dark matter (DM) consists primarily of bosons with mass $\ma\ll \eV$, then the DM density near the Sun guarantees that the number density of particles is large enough to be treated as an oscillating classical field.
The wavelike nature of such ultralight scalar fields---often referred to as axions---can lead to distinctive signatures that impact experimental searches.
In particular, interference among these waves results in a locally stochastic field that can potentially enhance or suppress the signal recorded by a laboratory experiment~\cite{Derevianko:2016vpm, Foster:2017hbq}.
In this work, we present a complete treatment of the stochastic properties of the gradient of the axion field, and explore its experimental consequences.

Axions are common in many extensions of the Standard Model, and correspond to Goldstone bosons that are produced when a global symmetry is broken.
The QCD axion remains one of the most compelling examples as it both provides a plausible DM candidate and resolves the strong CP problem~\cite{Peccei:1977hh,Peccei:1977ur,Weinberg:1977ma,Wilczek:1977pj,Kim:1979if,Shifman:1979if,Zhitnitsky:1980tq,Dine:1981rt,Preskill:1982cy,Abbott:1982af,Dine:1982ah}.
For the case of the QCD axion, there is a direct relationship between its mass and the symmetry-breaking scale in the theory.
More generally, axion-like particles (ALPs) can exist with a mass that is not set entirely by the symmetry-breaking scale.
Such ALPs can be copiously produced in string theories, for example~\cite{Arvanitaki:2009fg}.
Throughout this paper, we will use the word `axion' to refer to any ultralight scalar boson that couples to the axial current.  

Axions can couple to the Standard Model, opening the possibility of discovering them in the laboratory or beyond.  The axion coupling to fermions takes the form of a derivative term in the Lagrangian of the form $\mathcal{L} \propto g_{aff} \partial_\mu a \bar{f} \gamma^5 \gamma^\mu f$, which in the non-relativistic limit becomes $\mathcal{L} \propto g_{aff} \grada \cdot \vec{S}_f$ with $\vec{S}_f$ being the fermion spin.
Searches for these gradient interactions typically rely on sophisticated atomic-molecular-optical (AMO) or nuclear magnetic resonance (NMR) techniques, which are sensitive to small changes in, e.g., neutron spins~\cite{Graham:2013gfa,Graham:2017ivz,Aybas:2021cdk}.  DM searches for the axion-fermion couplings have been performed using data from neutron electric dipole moment (nEDM) experiments~\cite{Abel:2017rtm}, CASPEr~\cite{Garcon:2019inh,Wu:2019exd}, the E\"ot-Wash spin-polarized torsion balance~\cite{Terrano:2019clh}, and various atomic magnetometers~\cite{Kornack:2005,Vasilakis:2011,Brown:2011,Bloch:2019lcy}, including the NASDUCK experiment~\cite{Bloch:2021vnn}. Additional laboratory constraints come from searches for new long-range forces~\cite{Adelberger:2006dh,Terrano:2015sna} and invisible meson decays~\cite{Essig:2010gu}. The strongest constraints on the axion-fermion couplings are astrophysical, coming from axions produced in the Sun~\cite{Akerib:2017uem}, as well as neutron star~\cite{Beznogov:2018fda}, supernova~\cite{Carenza:2019pxu}, and white dwarf~\cite{Bertolami:2014wua} cooling.  These astrophysical constraints may however be subject to large uncertainties (see, e.g.,~Ref.~\cite{Bar:2019ifz}).

The pseudoscalar axion can also couple to photons through the Lagrangian operator  $\mathcal{L} \propto \frac{1}{4} g_{a\gamma\gamma} a F \widetilde{F}$, which in the non-relativistic limit reduces to $\mathcal{L} \propto g_{a\gamma\gamma}  a \vec{E} \cdot \vec{B}$.  This coupling has been the target of axion DM searches with experiments such as ADMX~\cite{Asztalos:2001tf,Du:2018uak,Braine:2019fqb}, HAYSTAC~\cite{Brubaker:2016ktl,Zhong:2018rsr}, ABRACADABRA~\cite{Kahn:2016aff,Ouellet:2018beu,Salemi:2021gck}, and SHAFT~\cite{Gramolin:2020ict}.
Additional constraints on $g_{a\gamma\gamma}$ that do not rely on the local axion DM density come from direct laboratory searches~\cite{Ehret:2010mh,Betz:2013dza,Ballou:2015cka}, axion helioscopes such as CAST~\cite{Anastassopoulos:2017ftl}, and other indirect astrophysical searches~\cite{Abramowski:2013oea,Ayala:2014pea,Payez:2014xsa,TheFermi-LAT:2016zue,Reynolds:2019uqt}.

\begin{figure*}
  \centering
  \includegraphics[width=\textwidth]{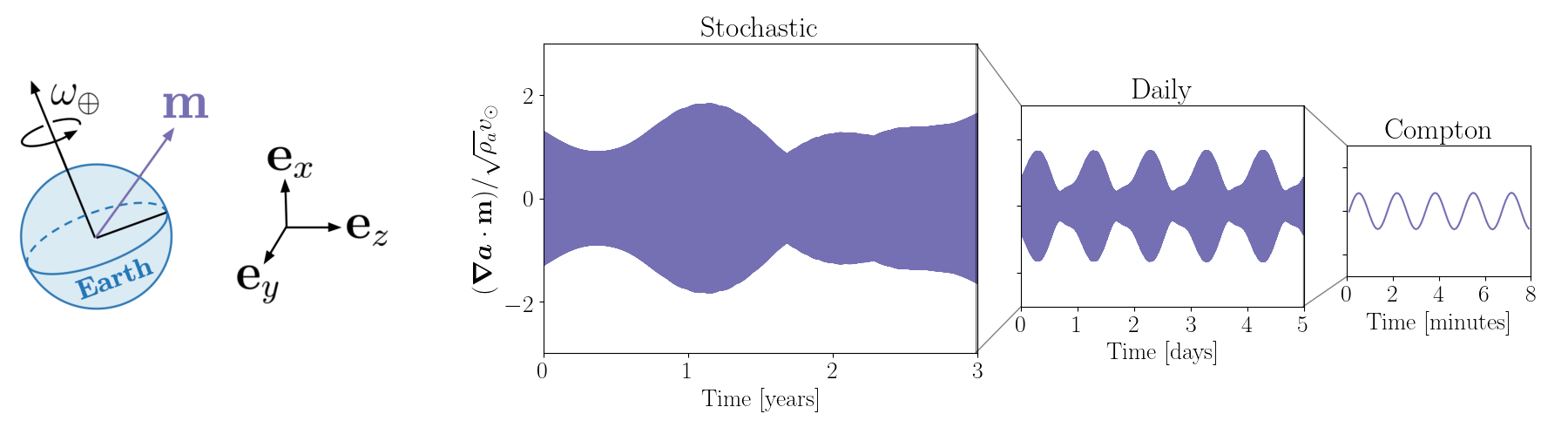}
  \caption{
    \textbf{Left:}
    A schematic illustration of the orientation of the laboratory measurement axis $\hat{\axis}$, which we assume has a fixed orientation relative to the surface of the Earth at some latitude and longitude, and our Galactic-frame basis $\{\hat{e}_x, \hat{e}_y, \hat{e}_z \}$.
    As the Earth rotates about its axis relative to the inertial Galactic frame, the Galactic-frame components of the measurement axis $\hat{\axis}$ oscillate with period of 1~day, thereby inducing daily modulation into the signal $\grada \cdot \hat{\axis}$.
    Note that the daily modulation of $\grada \cdot \hat{\axis}$ is periodic but not necessarily sinusoidal and that its exact form is stochastic, since it depends on the instantaneous direction of the axion gradient.
    \textbf{Right:}
    A single randomly generated time series for an axion gradient signal $\grada \cdot \hat{\axis}$ with mass $\ma/2\pi = 0.01~\mathrm{Hz}$ over the course of 3~years.
    On the smallest timescales, the signal undergoes rapid coherent oscillations with period $2\pi/\ma = 100~\mathrm{s}$.
    The amplitude of these oscillations modulates with period of $1~\mathrm{day}$ due to the rotation of the Earth, and evolves randomly on long timescales due to the coherence time $\tau_c \sim 6~\mathrm{months}$.
    \label{fig:cartoon}
    }
\end{figure*}

Any experiment searching for axion DM must contend with the unique phenomenology that arises from its wave-like behavior.  At a fixed position in space, the axion field oscillates with frequency $\omega \approx \ma + \frac{1}{2}\ma v^2$, where $v$ is the DM velocity. If the DM is virialized in the Milky Way with a velocity dispersion $\sigmav \sim 10^{-3}$, the observable classical field will be the superposition of many oscillating fields with a frequency dispersion $\Delta \omega \sim \frac{1}{2}\ma\sigmav^2$. The effect of this frequency dispersion is to introduce a coherence timescale, $\tau_c \sim 1/\ma\sigmav^2$, to the oscillations. On timescales $\tau \ll \tau_c$, the DM classical field is well-described by oscillations at the Compton frequency $\ma$ with a single amplitude and phase; however, on timescales $\tau \gtrsim \tau_c$, interference effects between the different frequency modes cause the amplitude and phase of oscillations to vary stochastically.   One consequence of this is the presence of localized regions where total destructive interference drives the axion field to zero, resulting in vortices~\cite{Hui:2020hbq}.

For experiments sensitive to the gradient of the axion field, there is an additional source of time dependence. This arises because such experiments do not measure the entire vector $\grada$, but the projection onto a sensitive axis, which is fixed in the laboratory frame.\footnote{It is possible for an experiment to have multiple sensitive axes. Although we do not consider this case specifically in this paper, the procedure outlined here is straightforward to generalize. See \Ref{Bloch:2021vnn} for an example.} Because the direction of $\grada$ is uniquely determined in the Galactic frame, this introduces an additional modulation of the signal due to the rotation of the Earth, as illustrated in the far left of \Fig{fig:cartoon}.
From an experimental perspective, this daily modulation is particularly useful when the Compton period is longer than 1 day, as it up-modulates the signal to a manageable frequency.

An axion gradient signal is therefore characterized by three timescales: the Compton period, $2\pi/\ma$; the coherence time, $\tau_c\sim 10^6/\ma$; and the period of the Earth's rotation, 1 day. The relative size of these timescales and the length of experimental observation can dramatically affect the phenomenology of the signal.
Fig.~\ref{fig:cartoon} provides a visualization of each of these timescales for an experiment that is sensitive to the axion field gradient. The figure shows the measured time series over the span of 3~years for an axion of mass $\ma/2\pi = 0.01$~Hz. For the particular axion mass shown in the figure, the three timescales obey the following hierarchy: $2\pi/\ma \ll 1~\mathrm{day} \ll \tau_c$. On very short timescales (right panel), the signal oscillates coherently with frequency $\ma$. The amplitude of these coherent oscillations undergoes daily modulations with a period of 1~day (middle panel).  On very long timescales (left panel), the signal decoheres and the overall amplitude fluctuates stochastically.

The stochastic nature of the axion field can have a profound effect on experimental observations---potentially leading to a suppression or enhancement in the signal, depending on the nature of the interference and its time dependence.  These effects must be properly accounted for in a complete data analysis to obtain reliable limits or properly recover a signal.
Although many experimental analyses have simply ignored the stochastic nature of the axion field, such effects have been well understood for experiments that are sensitive to the amplitude of the axion field itself~\cite{Derevianko:2016vpm,Foster:2017hbq,Centers:2019dyn}. However, these results do not immediately generalize to the case of the gradient of the axion field, where the \textit{direction} and not just the \textit{amplitude} of the axion field fluctuates stochastically.
Although there have been some attempts in recent literature to address the stochastic nature of the gradient of the axion field~\cite{Centers:2019dyn,Bloch:2019lcy}, they have relied on inadequate assumptions that do not capture the parametric freedom of an axion gradient signal.

In this paper, we present a general stochastic description for the gradient of the axion field as well as provide a likelihood formalism for the statistical analysis of experimental data.
This stochastic formalism can be used to analyze axion signals of any frequency, regardless of the coherence time.
We demonstrate, contrary to the assumptions of Ref.~\cite{Centers:2019dyn}, that the gradient of the axion field has both a random amplitude and direction, and that an analysis that ignores this effect has a non-negligible chance of failing to correctly identify an axion signal.
Sec.~\ref{sec:basic_formalism} introduces the basic formalism we use to model the superposition of axion states.  Then, in Sec.~\ref{sec:time_domain}, we demonstrate how to perform an analysis for data taken in the time domain.  Sec.~\ref{sec:coherent} specializes to the specific case of a coherent axion signal and summarizes the main conclusions of our work using some examples on mock data.  We conclude in Sec.~\ref{sec:conclusions}.  Three appendices are also included, which discuss how to derive the Central Limit Theorem (Appendix~\ref{app:clt}), how to estimate uncertainties using Ordinary Least Squares (Appendix~\ref{app:ols}), and how to evaluate the two-point correlation functions presented in the main text for the case of a Maxwellian velocity distribution (Appendix~\ref{app:max}).

\section{Superposition of Axion Waves}
\label{sec:basic_formalism}

This section presents the basic formalism for axion wave superposition.  We generalize the treatment of the axion field presented in \Ref{Foster:2017hbq} to the case of the axion field gradient. The formalism introduced here will be applied in Sec.~\ref{sec:time_domain} to construct a likelihood to analyze time-domain data.

In general, due to the time-dependent motion of the terrestrial laboratory relative to the Galactic frame, it is best to work with Galactic-frame velocities, $\vec{w}$, rather than laboratory-frame velocities, $\vec{v}$, as the underlying state variables. This transformation is given by the Galilean boost $\vec{v}(t) = \vec{w} + \vobsvec(t)$, where
\be
\vobsvec(t) = \vsunvec + \vearthvec(t) + \vlabvec(t)  
\label{eq:vobs}
\ee
is the velocity of the terrestrial laboratory relative to the Galactic frame, which depends on the Sun's velocity relative to the Galactic Center, $\vsunvec$, the Earth's velocity relative to the Sun, $\vearthvec(t)$, and the experiment's velocity relative to the Earth's center, $\vlabvec(t)$. 

We treat the axion as a classical field of mass $\ma$ consisting of a very large number of states, $N_a\gg 1$, and assume that the local axion density, $\rhodm$, comprises the entirety of the local DM. The field contribution from an individual state with Galactic-frame velocity $\vec{w}_{\lambda}$ is therefore
\be
a_\lambda(\vec{x}, t) = \frac{\sqrt{2\rhodm/N_a}}{\ma} \cos\left(E_\lambda(t) t + \vec{k}_\lambda(t) \cdot\vec{x} + \phi_\lambda \right) ,
\label{eq:state_field}
\ee
where $\lambda \in 1,2,...,N_a$ is an index that runs over all states, $E_\lambda(t) \approx \ma + \frac{1}{2}\ma v_\lambda(t)^2$ is the energy of the state, $\vec{k}_\lambda(t) \approx \ma\vec{v}_\lambda(t)$ is the momentum of the state, $\vec{v}_\lambda(t) = \vec{w}_\lambda + \vobsvec(t)$ is the laboratory-frame velocity of the state, and $\phi_\lambda \in[0,2\pi)$ is the phase of the state.
  The factor of $1/\sqrt{N_a}$ in the amplitude ensures that the total axion field, which is the superposition of all states, has root mean square density $\rhodm$.
  
The gradient of an individual axion state at a fixed point in space is therefore
\be
\grada_\lambda(t) = \sqrt{\frac{2\rhodm}{N_a}} \left( \vec{w}_\lambda + \vobsvec(t)\right)\cos\left(E_\lambda(t) t + \phi_\lambda \right) .
\label{eq:grad_ai}
\ee
Moving forward, the spatial dependence will be ignored because we only consider experiments at a fixed spatial location.

The total gradient of the axion field is obtained by summing \Eq{eq:grad_ai} over all individual states
\be
\grada(t) = \sum_\lambda \sqrt{\frac{2\rhodm}{N_a}} \left(\vec{w}_\lambda + \vobsvec(t)\right) \cos\left(E_\lambda(t) t + \phi_\lambda\right) , \el
\label{eq:grada_sum}
\ee
where each $\phi_\lambda$ is independent and identically distributed (IID) from the uniform distribution on the interval $[0,2\pi)$ and each $\vec{w}_{\lambda}$ is IID from the Galactic-frame velocity distribution $f(\vec{w})$.\footnote{Note that this construction assumes that the Galactic-frame DM velocity distribution is time-independent. If not, then \Eq{eq:grada_sum} would no longer be valid for all times, but only at a particular moment of time.
Fortunately, the dynamical timescale for the Milky Way is $\sim 300~\mathrm{Myr}$, which is much longer than the observation time of any human-timescale experiment, and the time dependence of the Galactic-frame velocity distribution function can be safely ignored.
}

Because the number of terms in the summation of \Eq{eq:grada_sum} is very large, and each term is an IID random variable, the Central Limit Theorem guarantees that $\grada(t)$ is a Gaussian process. That is, $\grada(t)$ is a normally distributed random vector at any particular time with a random time evolution obeying a particular two-point correlation function (see Appendix~\ref{app:clt}).

Experiments that are sensitive to the gradient of the axion field typically have a single axis of sensitivity. That is, they do not measure the vector field $\grada(t)$ but rather its projection onto some measurement axis, $\hat{\axis}(t)$. We assume that this axis is fixed in the laboratory frame, and therefore that it is time-dependent in the Galactic frame due to the rotation of the Earth about its axis, as illustrated in ~\Fig{fig:cartoon}. Because the direction of the vector field $\grada(t)$ is determined by the velocity of the axion particles, which are inertial in the Galactic frame, this projection causes the signal to undergo daily modulation with frequency $\omegaE=2\pi/\left(\mathrm{sidereal\ day}\right)$. The signal $\grada(t)\cdot\hat{\axis}(t)$ can be decomposed in the Galactic frame by writing
\be
\grada(t)\cdot\hat{\axis}(t) = \sum_{i=x,y,z}\left[ \grada(t)\cdot\hat{e}_i \right] \axis_i(t) \, ,
\label{eq:grad_dot_m}
\ee
where $\{ \hat{e}_x, \hat{e}_y, \hat{e}_z \}$ are an arbitrary set of orthonormal basis vectors fixed in the Galactic frame.  The components of $\hat{\axis}(t)$ in this basis, $\axis_i(t)=\hat{e}_i\cdot\hat{\axis}(t)$, undergo daily modulation due to the rotation of the Earth. For an axis that is fixed in the laboratory frame, this modulation can be parametrized as
\be
\axis_i(t) = C_i\cos\left(\omegaE t\right) + D_i\sin\left(\omegaE t\right) + E_i \, ,
\label{eq:modulation}
\ee
for $i=x,y,z$.  Here, $C_i, D_i, E_i$ are constants that depend on the position and orientation of $\hat{\axis}$ relative to the Earth's rotational axis.
For concreteness in the rest of this paper, we take $\hat{\axis}(t)$ to be the upward direction at $40\degree$~N latitude and $75\degree$~W longitude, and we always measure time from J2000.

\section{Analysis Strategy}
\label{sec:time_domain}

Next, we present a procedure for analyzing an observable signal in the time domain, using the construction of the axion field gradient in \Eq{eq:grada_sum}.  We begin with a description of the likelihood formalism in Sec.~\ref{sec:likelihood} before working through the general time-binned stochastic analysis in Sec.~\ref{sec:time_domain_binned}. 

\subsection{Likelihood Formalism}
\label{sec:likelihood}

Consider a time series of $N$ data points $\boldsymbol{D} = \{ D(t_n)\  |\ n\in1,...,N\}$, where each observation is the sum of a signal and background  contribution.  To write down a likelihood function for the observed data set, 
we must understand the probability distributions for both the signal and the background, which are assumed to be independent of each other.

The signal is related to the gradient of the axion field, 
\be
S(t_n) = g_{\rm eff} \grada(t_n)\cdot\hat{\axis}(t_n) \, ,
\label{eq:signal}
\ee
where $g_{\rm eff}$ is the effective axion coupling constant that is proportional to $g_{aff}$, but depends on the particular experiment at hand. For example, for a Helium-Potassium comagnetometer and an axion that couples only to neutrons, the observable can be taken to have dimensions of magnetic field with $g_{\rm eff} = g_{aNN}/\gamma_n$, where $\gamma_n$ is the gyromagnetic ratio of the neutron~\cite{Bloch:2019lcy}.  As discussed in \Sec{sec:basic_formalism}, the Central Limit Theorem guarantees that the components of $\grada(t)$ are jointly-distributed Gaussian random variables and thus, from \Eq{eq:signal}, that $S(t_n)$ forms a Gaussian process.  The uniform random phases guarantee that the mean of this process is always zero.
Therefore, the likelihood of the signal (ignoring background) is an $N$-dimensional zero-mean Gaussian function.

Throughout this work, the background is modeled by stationary Gaussian white noise with zero mean, although the generalization to other background probability distributions is straightforward.
Taking this together with the signal model, the total likelihood is 
\be
\mathcal{L}\left(\boldsymbol{D}\,|\,\boldsymbol{\theta}_{\rm sig}, \boldsymbol{\theta}_{\rm bkg}\right) = \frac{1}{\sqrt{(2\pi)^{6N}\mathrm{det}\boldsymbol{\Sigma}}} e^{-\frac{1}{2}\boldsymbol{D}^\intercal\boldsymbol{\Sigma}^{-1}\boldsymbol{D}} \, ,
\label{eq:full_likelihood}
\ee
where $\boldsymbol{\Sigma} = \boldsymbol{\Sigma}_{\rm sig}(\boldsymbol{\theta}_{\rm sig}) + \boldsymbol{\Sigma}_{\rm bkg}(\boldsymbol{\theta}_{\rm bkg})$ is the combined covariance matrix for the signal and background and $\boldsymbol{\theta}_{\rm sig}$ and $\boldsymbol{\theta}_{\rm bkg}$ are the signal and background model parameters, respectively.  For the white noise scenario, the background covariance matrix is proportional to the identity matrix, $\boldsymbol{\Sigma}_\text{bkg} = \sigma_{\rm bkg}^2 \mathbf{I}$, where $\sigma_{\rm bkg}$ is the root-mean-square (RMS) noise. The signal model parameters are the axion mass, $\ma$, and the coupling constant, $g_{\rm eff}$.  

The final ingredient for defining the likelihood function is the covariance matrix, or two-point correlator, of the signal:
\be
\Sigma_{{\rm sig},nm} = \langle S(t_n) S(t_m)\rangle \, .
\ee
Although in principle it is possible to compute this object, the result is not particularly useful because the $N\times N$ covariance matrix is unwieldy with $N$ typically being extremely large.  However, we will show below that the effective size of the data set can be significantly reduced and that a computationally feasible analysis in the time domain is always possible. 

The analysis could also be done in the frequency domain, which may be more efficient if a Fast Fourier Transform can be performed on the data and the integration time is much longer than the coherence time, $\texp \gg \tau_c$. However, the daily modulation of the axion gradient signal, which leads to three peaks in frequency space at $m_a$ and $m_a \pm \omega_\oplus$, adds significant complication to a frequency domain analysis compared to an analysis of a signal proportional to the scalar axion field~\cite{Foster:2017hbq}. We will present an example of a full frequency domain analysis in upcoming work (see also \Ref{Bloch:2021vnn}).

\subsection{Time-Binned Stochastic Analysis}
\label{sec:time_domain_binned}
Next, we discuss an efficient manner of performing a time-binned analysis for an axion signal with a relatively short coherence time compared to the experimental integration time, $\tau_c \lesssim T_{\rm exp}$.  In this regime, the signal is not coherent over the lifetime of the experiment and it is necessary to fully characterize the stochastic effects of the axion field.  To begin, the expression for the signal in \Eq{eq:signal} is rewritten to separate out the deterministic and stochastic time dependencies.  Specifically, 
\be
S(t) &=& \sum_{i=x,y,z} \left[A_i(t) \cos(\ma t) - B_i(t) \sin(\ma t)\right] \axis_i(t) \, , \el
\label{eq:signal_decomposition}
\ee
where
\be
A_i(t) &=& g_{\rm eff}\sqrt{\frac{2\rhodm}{N_a}} \sum_{\lambda} v_{\lambda,i}(t)\cos\left(\frac{1}{2}\ma v_\lambda(t)^2t+\phi_\lambda\right) \el
B_i(t) &=& g_{\rm eff}\sqrt{\frac{2\rhodm}{N_a}} \sum_{\lambda} v_{\lambda,i}(t)\sin\left(\frac{1}{2}\ma v_\lambda(t)^2t+\phi_\lambda\right) \el 
\label{eq:ABdef}
\ee
and $v_{\lambda, i}(t)$ is the $\hat{e}_i$ component of the vector $\vec{v}_\lambda(t)=\vec{w}_\lambda + \vec{v}_{\rm obs}(t)$.  The utility of this decomposition is immediately apparent.  The signal manifestly oscillates at the Compton frequency $\ma$, with daily modulations at frequency $\omegaE$ due to the $\axis_i(t)$ terms.  The $A_i(t)$ and $B_i(t)$ coefficients, which vary on the coherence timescale, are Gaussian random variables that encode the stochastic behavior of the axion field.

We now consider the behavior of the signal on small timescales.
We divide the time series $\mathbf{S} = \{ S(t_n)\  |\ n\in1,...,N\}$ into $P$ bins of size $\Delta t \ll 2\pi/\omegaE, \tau_c$ centered at times $t_p$.
Within each bin, the daily modulations and stochastic fluctuations can be neglected, so the signal in the $p^\text{th}$ bin can be written as a coherent oscillation
\be
S_p(t) = \tilde{A}_p \cos(\ma t) - \tilde{B}_p\sin(\ma t) \, ,
\label{eq:signal_binned}
\ee
where
\be
\tilde{A}_p &=& \sum_{i=x,y,z} \axis_i(t_p) A_i(t_p) \el
\tilde{B}_p &=& \sum_{i=x,y,z} \axis_i(t_p) B_i(t_p)
\ee
are constant within each bin and vary smoothly (and stochastically) in time from bin to bin.

Because $\tau_c \sim 10^6/\ma$ and we need only consider $\ma\gg \omegaE$,\footnote{In this subsection, we assume that the coherence time is relatively short, $\tau_c\lesssim \texp$. Assuming a human-timescale experiment with $\texp\lesssim 10\ \mathrm{yr}$, this implies that $\ma \gtrsim 40\ \omegaE$. When this is not satisfied, the analysis in \Sec{sec:coherent} can be used.} it is always possible to choose the bin size $\Delta t$ such that there are many coherent oscillations within each bin.
As a result, the coefficients $\tilde{A}_p$ and $\tilde{B}_p$ can be measured by fitting the data within each bin to the form of \Eq{eq:signal_binned} with, e.g., an ordinary least squares procedure.
This motivates an analysis strategy where the $2P$ coefficients $\tilde{A}_p$ and $\tilde{B}_p$ are analyzed instead of the full time series.
Because there are many coherent oscillations within each bin, this method achieves a significant reduction in the effective size of the data set.\footnote{The coefficients $\tilde{A}_p$ and $\tilde{B}_p$, and indeed the bin sizes themselves, have to be recomputed for each axion mass $\ma$; however, so does the inverse and determinant of the covariance matrix. Since the bandwidth of any experiment is not larger than $\mathcal{O}(N)$ frequency points, it will always be more efficient to recompute the ordinary least squares coefficients $\tilde{A}_p$ and $\tilde{B}_p$ than to work with the full $N\times N$ covariance matrix.}

We now consider the statistical analysis of such a compressed data set consisting of $P$ measurements of the two coefficients $\{\tilde{A}(t_p), \tilde{B}(t_p)\}$ at the binned times $t_p$ for  $p~=~1,\ldots,P$. \Eq{eq:ABdef} and the Central Limit Theorem guarantee that the time series of these coefficients form joint stationary Gaussian processes. Assuming Gaussian white noise, the uncertainty on the recovered ordinary least squares coefficients is also Gaussian distributed, but with uncertainty $ \tilde{\sigma}~\approx~\sigma_{\rm bkg}\sqrt{2P/N}$ (see Appendix~\ref{app:ols} for further details).
Therefore, a likelihood of the form in \Eq{eq:full_likelihood} is still valid, and we need only compute the two-point correlation functions in order to construct the covariance matrix:
\be
\langle \tilde{A}(t)\tilde{A}(t')\rangle &=& \sum_{i,j} \axis_i(t) m_j(t')\langle A_i(t)A_j(t')\rangle \el
\langle \tilde{B}(t)\tilde{B}(t')\rangle &=& \sum_{i,j} \axis_i(t) m_j(t')\langle B_i(t)B_j(t')\rangle \el
\langle \tilde{A}(t)\tilde{B}(t')\rangle &=& \sum_{i,j} \axis_i(t) m_j(t')\langle A_i(t)B_j(t')\rangle \, .
\label{eq:ABcorrelators}
\ee
The expectation values in \Eq{eq:ABcorrelators} can be evaluated using \Eq{eq:clt_cov}, which, after integrating over the random phases, gives
\be
\langle A_i(t)A_{j}(t') \rangle &=& g_{\rm eff}^2\rhodm \int \dd^3 \vec{w}\ v_i(t)v_j(t') f(\vec{w}) \cos\left(\Delta \varpi\right) \el
\langle A_i(t)B_{j}(t') \rangle &=& g_{\rm eff}^2\rhodm \int \dd^3\vec{w}\ v_i(t) v_j(t') f(\vec{w}) \sin\left(\Delta \varpi \right) \el \, 
\langle A_j(t)A_{j}(t')\rangle &=& \langle B_j(t)B_{j}(t')\rangle \, ,
\label{eq:time_domain_cov_integral}
\ee
where $\Delta \varpi = \frac{1}{2}\ma\left( v^2(t')t' - v^2(t)t\right)$ and $\vec{v}(t) = \vec{w} + \vec{v}_{\rm obs}(t)$.
This is the most general solution for the covariance matrix, and can be computed for any velocity distribution of the axion DM.

While the discussion until now has remained general, it is useful to consider the specific case where the Galactic-frame DM velocity distribution is Maxwellian,
\be
f(\vec{w}) = \frac{1}{\left(2\pi\sigmav^2\right)^{3/2}}e^{-w^2 /2\sigmav^2} \, ,
\label{eq:maxwell_boltzmann}
\ee
with velocity dispersion $\sigmav \approx 155\ \mathrm{km/s}$~\cite{Koposov:2010,Bovy:2012,Eilers:2019}. Note that to compute the integrals in \Eq{eq:time_domain_cov_integral} analytically, we do not introduce a cutoff of the distribution at the escape velocity.

Additionally, we make the simplifying assumption that $\vobsvec(t) \approx \vsunvec$, ignoring the (comparatively small) time-dependent contributions of the Earth's velocity around the Sun, and the laboratory's velocity around the Earth's center. We take $\vsunvec = (11, 232, 7)~\mathrm{km/s}$ in the standard Galactic $(U,V,W)$ coordinate basis\footnote{The $(U,V,W)$ coordinate basis is aligned such that $\hat{e}_U$ is the direction from the Solar System barycenter towards the Galactic center, $\hat{e}_V$ is the direction of the local Milky Way disk rotation, and $\hat{e}_W$ points towards the Galactic North Pole.}~\cite{Schonrich:2010}. This assumption allows the integrals in \Eq{eq:time_domain_cov_integral} to be computed analytically via a change of variables from $\vec{w} \to \vec{v} = \vec{w} + \vsunvec$. We find that the two-point correlation functions evaluated using this approximation differ from the full solution by less than 10\%.

\begin{figure*}[t]
  \centering
  \includegraphics[width=\textwidth]{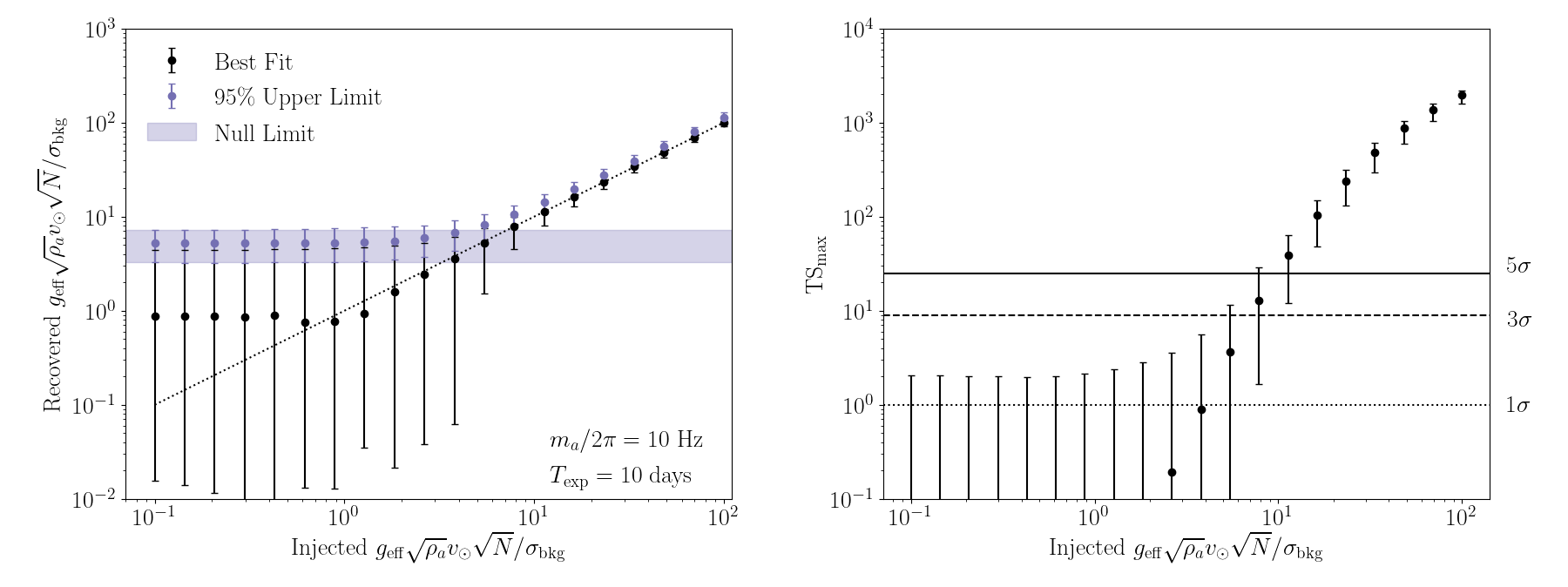}
  \caption{
    \textbf{Left:}
    Signal injection plot using mock data to validate the analysis procedure outlined in \Sec{sec:time_domain_binned}. The mock data were generated with a sampling rate of $50~\mathrm{Hz}$ for $10~\mathrm{days}$ with stationary Gaussian white noise background with standard deviation $\sigma_{\rm bkg}=10^{-8}\sqrt{\rhodm}\vsun\ \mathrm{GeV}^{-1}$. The signal was generated with an axion mass of $\ma/2\pi=10~\mathrm{Hz}$ for 20 different couplings $g_{\rm eff}$, shown on the horizontal axis. For each value of injected coupling, we simulate 100 random mock data sets, and analyze each data set using the Gaussian likelihood procedure described in the text. The black (\blueish) points indicate the median best-fit (95\% upper limit) couplings over the 100 mock data sets as a function of the injected signal strength. The horizontal band indicates the 95\% upper limit with zero injected signal.
    \textbf{Right:}
    The median discovery test statistic, $\ts$, obtained in the analysis of the mock data as a function of the injected signal strength. The horizontal lines indicate 1-, 3-, and 5-$\sigma$ local significance. In both panels, the error bars indicate the 5$^{\rm th}$ and 95$^{\rm th}$ percentiles of the best-fit coupling, recovered limit, or test statistic. The couplings are reported as a dimensionless signal-to-noise ratio $g_{\rm eff}\sqrt{\rhodm}\vsun\sqrt{N}/\sigma_{\rm bkg}$, where $\sigma_{\rm bkg}/\sqrt{N}$ is the relative scaling of the background for an infinitely coherent signal.
\label{fig:full_analysis_signal_injection}
      }
\end{figure*}

Because $f(\vec{w})$ is an isotropic function and $\vsunvec$ is constant, it is useful to align our otherwise arbitrary basis vectors $\{\hat{e}_x, \hat{e}_y,\hat{e}_z\}$ with the unique direction $\vsunvec$. Specifically, we choose $\hat{e}_z\parallel\vsunvec$. This guarantees that all two-point correlation functions are proportional to $\delta_{ij}$, i.e., the $x$, $y$, and $z$ components are statistically independent of one another, and that there is an $x\leftrightarrow y$ symmetry.
Substituting \Eq{eq:maxwell_boltzmann} into \Eq{eq:time_domain_cov_integral}, we find that 
\be
\langle A_z(t)A_z(t')\rangle &=& \mathcal{A}_{\parallel}(\xi)\cos\Psi_{\parallel}(\xi) \el
\langle A_z(t)B_z(t')\rangle &=& \mathcal{A}_{\parallel}(\xi)\sin\Psi_{\parallel}(\xi) \el
\langle A_x(t)A_x(t')\rangle &=& \mathcal{A}_{\perp}(\xi)\cos\Psi_{\perp}(\xi) \el
\langle A_x(t)B_x(t')\rangle &=& \mathcal{A}_{\perp}(\xi)\sin\Psi_{\perp}(\xi) \, ,
\label{eq:ABcorr}
\ee
where we have defined $\xi \equiv \ma\sigmav^2(t'-t)$ and the amplitudes and phases are:
\begin{widetext}
\begin{equation}
\begin{split}
\mathcal{A}_{\parallel}(\xi) &= \frac{g_{\rm eff}^2\rhodm}{(1+\xi^2)^{7/4}}\exp\left[-\frac{\vsun^2}{2\sigmav^2}\frac{\xi^2}{(1+\xi^2)}\right]\sqrt{(\sigmav^2+\vsun^2)^2+\sigmav^4 \xi^2} \\
\Psi_{\parallel}(\xi) &= \frac{\vsun^2}{2\sigmav^2}\frac{\xi}{(1+\xi^2)} + \frac{7}{2}\arctan(\xi) - \arctan\left(\frac{\sigmav^2 \xi}{\sigmav^2+\vsun^2}\right) 
\end{split}
\quad\quad
\begin{split}
\mathcal{A}_{\perp}(\xi) &= \frac{g_{\rm eff}^2\rhodm\sigmav^2}{(1+\xi^2)^{5/4}}\exp\left[-\frac{\vsun^2}{2\sigmav^2}\frac{\xi^2}{(1+\xi^2)}\right] \\
\Psi_{\perp}(\xi) &= \frac{\vsun^2}{2\sigmav^2}\frac{\xi}{(1+\xi^2)} + \frac{5}{2}\arctan(\xi) \, .
\end{split}
\label{eq:amplitude_phase}
\end{equation}
\end{widetext}
Further details on the evaluation of these correlation functions are provided in Appendix~\ref{app:max}.

We validate this likelihood and analysis procedure by generating 10 days of mock data sampled at $50~\mathrm{Hz}$ with Gaussian white noise and injected signal with an axion mass of $\ma/2\pi = 10~\mathrm{Hz}$ and 20 different values of the coupling $g_{\rm eff}$. We repeat this for 100 different random iterations of noise and axion signal. For each mock data set, we bin the data in bins of length $\Delta t = 10^5/\ma$\footnote{Recall that, in general, the bin size must satisfy $\Delta t \ll \tau_c, 1~\mathrm{day}$. In practice, we find that choosing $\Delta t = \mathrm{min}(10^5/\ma, 2.4~\mathrm{hr})$ is enough to ensure that the essential features appear in \Fig{fig:full_analysis_signal_injection}.} and obtain the $\tilde{A}$ and $\tilde{B}$ coefficients in each bin from ordinary least squares fitting.

For the likelihood procedure, we use a multivariate Gaussian likelihood of the form in \Eq{eq:full_likelihood}, but with the covariance matrix given by the two-point correlation functions in \Eq{eq:ABcorrelators}, which are computed using \Eq{eq:ABcorr}.  The uncertainty of the $\tilde{A}$ and $\tilde{B}$ coefficients, $\tilde{\sigma}$, is treated as a nuisance parameter and accounted for using the profile likelihood method~\cite{Rolke:2004mj, Cowan:2010js}.  The best-fit coupling for a given axion mass, $g_{\rm best}$, is obtained by maximizing the likelihood $\mathcal{L} \left(\mathbf{D} \, | \, m_a, g_{\rm eff}\right)$ for data set $\mathbf{D}$.  The test statistic is defined as 
\begin{equation}
{\rm TS} \equiv  2 \left[ \log \mathcal{L}\left(\mathbf{D} \, | m_a,  \, g_{\rm best} \right) - \log \mathcal{L}\left(\mathbf{D} \, | \, m_a, g_{\rm eff}\right) \right] \, ,
\label{eq:ts}
\end{equation}
from which it follows that the 95\% upper confidence limit on the coupling corresponds to the value of $g_{\rm eff}$ where $\text{TS} = 2.71$.  To characterize the significance of a potential discovery of an axion with $g_{\rm best}$, we use $\ts$,  defined as in \Eq{eq:ts}, but where the maximized likelihood is compared to the null hypothesis with $g_{\rm eff} = 0$.

The results of our mock analysis are displayed in \Fig{fig:full_analysis_signal_injection}, and serve as a validation for the analysis procedure. The value of the true injected axion coupling is shown on the horizontal axis of both panels. In the left panel, the black (purple) points indicate the median best-fit (95\% upper limit) couplings recovered over the 100 random mock datasets for each value of injected coupling. The right panel shows the median discovery test-statistic obtained for each value of injected coupling. At small injected couplings, the discovery significance is low and the 95\% upper limit couplings asymptote to a constant value, as expected. At large injected couplings, the discovery significance increases dramatically, the best-fit coupling becomes centered on the true injected value, and the error bars decrease, as expected in the regime where the signal-to-noise ratio is large. In this regime, the 95\% upper limit is always above the best-fit coupling and above the true injected coupling 95\% of the time.

\section{The Coherent Limit}
\label{sec:coherent}

\begin{figure*}[ht] 
   \centering
   \includegraphics[width=\textwidth]{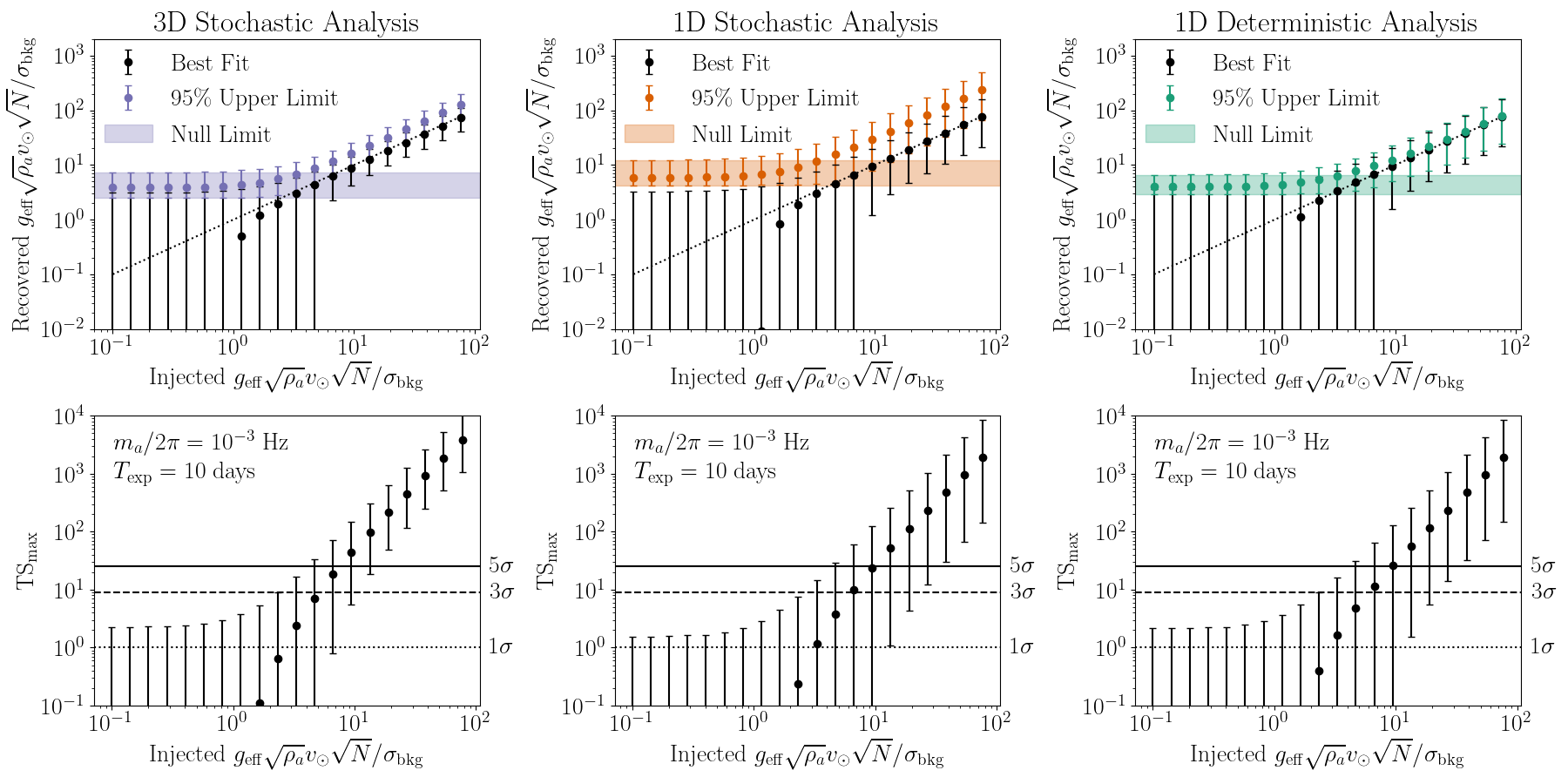}
   \caption{Signal injection plots using mock data to test the benchmark analysis procedures considered in \Sec{sec:coherent}. Each column is analogous to \Fig{fig:full_analysis_signal_injection} but for the 3D Stochastic (\textbf{Left}), 1D Stochastic (\textbf{Center}), and 1D Deterministic (\textbf{Right}) models. The same mock data were used for all three analyses. The mock data were generated with a sampling rate of $1~\mathrm{mHz}$ for $10~\mathrm{days}$ with Gaussian white noise background with standard deviation $\sigma_B=10^{-8}\sqrt{\rhodm}\vsun\ \mathrm{GeV}^{-1}$.  The signal was generated with an axion mass of $\ma/2\pi=1~\mathrm{mHz}$ for 20 different couplings $g_{\rm eff}$. For each value of injected coupling, we simulate 10,000 random mock data sets, and analyze each data set using the corresponding procedure, as described in the text.  The 1D~Stochastic and 1D~Deterministic models set an incorrect 95\% upper limit on $g_{\rm eff}$ and exhibit more variability in their $\ts$, as compared to the 3D~Stochastic model.  Additionally, for large injected coupling, the 95\% upper limit for the 1D~Deterministic model falls below the true value about 50\% time, when this should only occur about $5\%$ of the time for a correct model.
     \label{fig:signal_injection}
   }
\end{figure*}
This section focuses on the limit of very long axion coherence times, $\tau_c \gg \texp$, where the signal can be treated as coherent for the entire runtime of the experiment. For simplicity, we continue to assume a Maxwellian velocity distribution function and take $\vec{v}_{\rm obs} \approx \vec{v}_{\odot}$.
As seen in \Eqs{eq:ABcorr}{eq:amplitude_phase}, there is an analytic expression for the two-point correlations in this limit.   In this case, $\xi \rightarrow 0$ and we have $\mathcal{A}_{\parallel} = g_{\rm eff}^2 \rhodm \left(\sigmav^2+\vsun^2\right)$, $\mathcal{A}_{\perp}=g_{\rm eff}^2\rhodm\sigmav^2$, and $\Psi_{\parallel}=\Psi_{\perp}=0$. Therefore, all cross-correlations vanish and each auto-correlation matrix becomes singular.  This is because, in the limit of long coherence times, each coefficient $A_z(t)$, $B_z(t)$, ... does not vary over the course of the experiment and can be treated as a single Gaussian random variable, rather than a Gaussian process. This is consistent with our intuition that the coefficients should only vary on timescales of order the coherence time, and can be treated as constants on much shorter timescales. Taking these six Gaussian random variables that fully specify the signal and rescaling them so that they obey a standard normal distribution, we obtain the compact expression:
\be
S(t) &=&  g_{\rm eff} \sqrt{\rhodm \left(\sigmav^2+\vsun^2\right)}\alpha_z \cos(\ma t + \phi_z)\, \axis_z(t) \el
&& +  g_{\rm eff}\sqrt{\rhodm \sigmav^2}\alpha_y \cos(\ma t + \phi_y)\, \axis_y(t) \el
&& + g_{\rm eff}\sqrt{\rhodm \sigmav^2}\alpha_x \cos(\ma t + \phi_x)\, \axis_x(t) \, ,
\label{eq:coherent}
\ee
where $\alpha_i$ are three Rayleigh-distributed amplitudes, and $\phi_i$ are three uniformly-distributed random phases.\footnote{If $X$ and $Y$ are standard normal random variables, then $\sqrt{X^2+Y^2}$ follows a Rayleigh distribution, and $\arctan(Y/X)$ follows a uniform distribution.} We stress that this result is only valid for $\tau_c \gg \texp$ in a coordinate system defined such that $\hat{e}_z\parallel \vsunvec$ and for the Maxwellian velocity distribution.  Importantly, \Eq{eq:coherent} contains contributions from the overlap of all three Galactic basis vectors with the sensitive axis $\hat{\axis}$, and  each component in the Galactic basis has an independent random amplitude and phase.

  \begin{figure*}[t] 
    \centering
    \includegraphics[width=\textwidth]{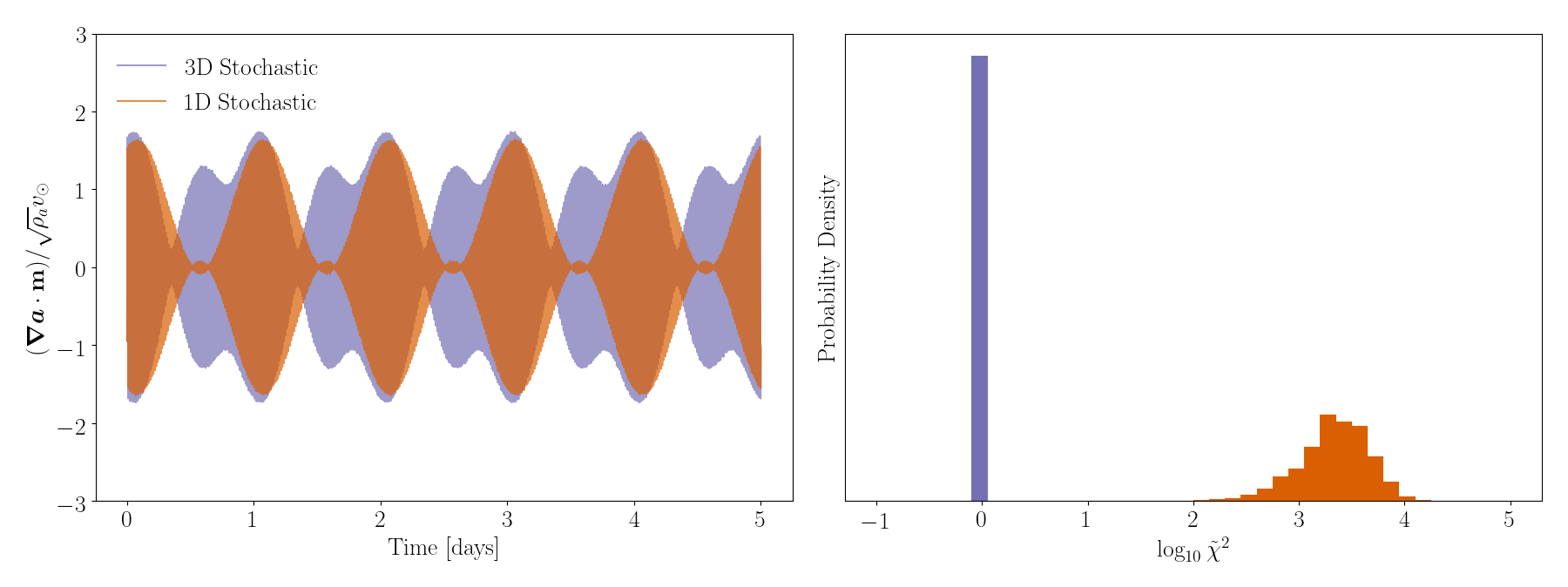}
    \caption{\textbf{Left:} Randomly generated time series for an axion gradient signal $\grada\cdot\hat{\axis}$ in the coherent limit. The data were generated with an axion mass of $\ma/2\pi = 10^{-3}~\mathrm{Hz}$ ($\tau_c \sim 5~\mathrm{yr}$) for 5 days. The data include Gaussian white noise with standard deviation $\sigma_{\rm bkg} = 10^{-2}\sqrt{\rhodm} \vsun$. The \blueish\ curve is a random realization of a true axion signal, the 3D Stochastic model, given by \Eq{eq:coherent} with a set of randomly generated amplitudes $\alpha_i$ and phases $\phi_i$. The \reddish\ curve is a realization of the 1D Stochastic model given in \Eq{eq:1d_stochastic} with the same amplitude and phase in the $\hat{e}_z$ component as in the realization of the 3D model, but no contribution from the $\hat{e}_x$ or $\hat{e}_y$ components. The 1D model is an insufficient approximation of the full 3D model. Note, in particular, that the amplitude modulation in the \reddish\ curve is determined entirely by $\axis_z(t)$, while in the \blueish\ curve it is determined by a random linear combination of $\axis_x(t)$, $\axis_y(t)$, and $\axis_z(t)$, giving the \blueish\ curve parametrically more freedom than the \reddish\ curve. \textbf{Right}: The distribution of the reduced chi-squared statistic for the correct 3D Stochastic model and the 1D Stochastic model over many random iterations of the same time series. For each model, the best-fit amplitudes and phases are obtained by fitting the mock data to either \Eq{eq:coherent} or \Eq{eq:1d_stochastic}. The reduced chi-squared statistic is defined as $\tilde{\chi}^2 = \sum_n \frac{S(t_n) - D_n}{\nu \sigma_{\rm bkg}}$, where $S(t_n)$ is the predicted signal for the model given the least-squares amplitudes and phases and $\nu$ is the number of degrees of freedom equal to $N$ minus the number of fit parameters. The reduced chi-squared statistic serves as a measure of goodness-of-fit for the models, with $\tilde{\chi}^2\approx1$ indicating a good fit, as is the case for the 3D model, and $\tilde{\chi}^2\gg 1$ indicating a poor fit, as is the case for the 1D model.
      \label{fig:components}
      }
  \end{figure*}

To analyze a data set in the coherent limit, we can simply fit the data to the form of \Eq{eq:coherent}, rather than constructing a multivariate Gaussian likelihood function with a covariance matrix that encapsulates the stochastic fluctuations in the signal.  Although the amplitudes and phases in \Eq{eq:coherent} are random variables, they can be treated as nuisance model parameters. The true likelihood would then be obtained in the usual way by marginalizing over the nuisance parameters.  Unfortunately, with six nuisance parameters, the marginalization integrals quickly become analytically intractable and numerically cumbersome, especially for relatively large data sets.
Fortunately, we can again take advantage of ordinary least squares fitting to dramatically reduce the effective size of the data set, thereby enabling a modified likelihood function to be used.

\Eq{eq:coherent} is equivalent to an ordinary least squares problem of the form
\be
S(t) = \sum_{i=x,y,z} \left[A_i\cos(\ma t) - B_i\sin(\ma t)\right] \axis_i(t) \, ,
\label{eq:coherent_fit}
\ee
and the data set can be reduced from a time series of size $N$ to the six measured least-squares coefficients $\left\{ (A_i, B_i) \, | \, i=x,y,z\right\}$.  Each measured coefficient comes with an associated uncertainty from the fitting procedure.  For Gaussian white noise with scale $\sigma_{\rm bkg}$, the resulting uncertainty on the coefficients is also Gaussian. Unlike in \Sec{sec:time_domain_binned}, however, this uncertainty is different for each coefficient. By symmetry, the uncertainty on each $A_i$ and $B_i$ should be the same, which we will denote $\tilde{\sigma}_i$, with each uncertainty scaling roughly as $\tilde{\sigma}_i \sim \sigma_{\rm bkg}/\sqrt{N}$, but with the exact form given by \Eq{eq:general_uncertainty}.
Since each $A_i$ and $B_i$ are independent Gaussian random variables, the total likelihood is the product of the individual probabilities of observing the given coefficients, marginalized over the underlying probability distributions. Because the underlying noise distribution is Gaussian as well, the result is a Gaussian likelihood
\be
\mathcal{L}(A_i, B_i, \tilde{\sigma}_{i}|g_{\rm eff}) &=& \prod_{i=x,y,z}\frac{e^{-(A_i^2 + B_i^2)/2\sigma_i^2(g_{\rm eff},\tilde{\sigma}_{i})}}{2\pi\sigma_i^2(g_{\rm eff},\tilde{\sigma}_{i})} \, ,
\label{eq:coherent_likelihood}
\ee
where
\be
\sigma_x^2(g_{\rm eff},\tilde{\sigma}_{x}) &=& g_{\rm eff}^2\rhodm \sigmav^2 + \tilde{\sigma}_{x}^2\el
\sigma_y^2(g_{\rm eff},\tilde{\sigma}_{y}) &=& g_{\rm eff}^2\rhodm \sigmav^2 + \tilde{\sigma}_{y}^2\el
\sigma_z^2(g_{\rm eff},\tilde{\sigma}_{z}) &=& g_{\rm eff}^2\rhodm (\sigmav^2 + \vsun^2) + \tilde{\sigma}_{z}^2 \, .
\ee

Note that there is a near-degeneracy between $g_{\rm eff}$ and $\sigma_{\rm bkg}$ (through each of the $\tilde{\sigma}_i$). In \Sec{sec:time_domain_binned}, we treated $\sigma_{\rm bkg}$ as unknown and marginalized over $\tilde{\sigma}$ as a nuisance parameter. However, due to the degeneracy between $g_{\rm eff}$ and $\sigma_{\rm bkg}$ in \Eq{eq:coherent_likelihood}, we cannot repeat the same procedure and must assume that $\sigma_{\rm bkg}$ is measured independently. For example, $\sigma_{\rm bkg}$ can be determined via $\sigma_{\rm bkg} \approx \sqrt{\mathrm{SSR}/N}$, where $\mathrm{SSR}$ is the minimum sum of the squares of the residuals obtained in the fitting procedure. We will therefore assume that $\sigma_{\rm bkg}$ and the fit coefficient uncertainties, $\tilde{\sigma}_{i}$, are known.

We validate this likelihood and analysis procedure by generating 10 days of mock data sampled at $10~\mathrm{mHz}$ with Gaussian white noise and injected signal with an axion frequency of $\ma/2\pi = 1~\mathrm{mHz}$ and 20 different values of the coupling $g_{\rm eff}$. We repeat this for $10^4$ different random iterations of noise and axion signal. For each mock data set, we obtain the six $A_i$ and $B_i$ coefficients as well as the uncertainty estimates $\tilde{\sigma}_{i}$ via ordinary least squares fitting.
We then compute the best-fit coupling $g_{\rm best}$, the discovery test statistic $\ts$, and the 95\% upper limit using the likelihood in \Eq{eq:coherent_likelihood}, following the same procedure discussed in \Sec{sec:time_domain_binned}.

The results of the mock analysis are displayed in the left column of \Fig{fig:signal_injection}, which is analogous to \Fig{fig:full_analysis_signal_injection}, but for the coherent analysis discussed above.  The discovery significance and 95\% upper limit couplings behave as desired across all injected couplings.  However, it is worth noting that the error bars on the recovered couplings and $\ts$ do not decrease in the limit of large injected coupling; this is expected behavior because, with much less than one coherence time of data, the stochastic fluctuations do not ``average out.''

Next, we compare the data model and analysis procedure outlined in this section against two benchmark models that do not fully encapsulate the behavior of \Eq{eq:coherent}. Our goal is to motivate the use of the full stochastic axion model presented in this paper and show that certain simplifying assumptions that have previously been made in the literature can fail to capture the crucial phenomenology of a true underlying axion signal. This can result in unreliable upper limits from experimental data and, more dramatically, missed discoveries of axion signals.

The first benchmark model we consider is purely deterministic and equivalent to the full stochastic model in \Eq{eq:coherent} in the limit of zero velocity dispersion. In this case, the direction of $\grada$ is the same as the average DM velocity $\vobsvec(t)\approx \vsunvec$, and therefore, the signal is proportional to $\vsunvec \cdot \hat{\axis}(t) = \vsun \axis_z(t)$: 
\be
S_{\rm 1D, det.}(t) = g_{\rm eff} \sqrt{2\rhodm \vsun^2} \cos(\ma t+\phi) \, \axis_z(t) \,.
\label{eq:deterministic}
\ee
Here, $\phi$ is a random phase, but note that the amplitude of oscillations is determined solely by the axion coupling $g_{\rm eff}$ and astrophysical parameters. The factor of $\sqrt{2}$ in the amplitude ensures that the signal has the same RMS as \Eq{eq:coherent} in the limit of zero velocity dispersion.
We refer to this model as the 1D Deterministic model because it assumes that the axion gradient is always in one direction and is fully deterministic.

The second benchmark model is inspired by the treatment of the gradient of the axion field in Ref.~\cite{Centers:2019dyn}. This model again assumes that the direction of $\grada$ is determined solely by $\vsunvec$, so that the signal is still proportional to $\vsun\axis_z(t)$; however, this model includes an additional random amplitude to account for stochastic interference effects:
\be
S_{\rm 1D, stoch.}(t) = g_{\rm eff} \sqrt{\rhodm \vsun^2} \alpha \cos(\ma t+\phi) \, \axis_z(t) \,.
\label{eq:1d_stochastic}
\ee
Here, $\alpha$ is a Rayleigh-distributed random amplitude, and $\phi$ is a uniformly-distributed random phase. We refer to this model as the 1D Stochastic model.\footnote{Note that the 1D Stochastic model lacks self-consistency because there are no stochastic amplitude fluctuations in the limit of zero velocity dispersion.}  For the sake of comparison, we refer to the correct model given in \Eq{eq:coherent} as the 3D Stochastic model, since it does not fix the direction of $\grada$, but assigns a random amplitude and phase independently to each of the three components of $\grada$, effectively giving the axion gradient a random direction as well as a random amplitude.

The 1D Stochastic model is motivated by previous studies, such as  Refs.~\cite{Centers:2019dyn} and~\cite{Bloch:2019lcy}, that have incorrectly assumed that the axion signal contains a single amplitude and phase coming from the overlap of the sensitive axis with the direction of the average DM velocity, $\hat{e}_z$.   As illustrated in \Fig{fig:components}, this incorrect assumption implies a dramatic difference in the behavior of the signal time series.  In the left panel of \Fig{fig:components}, the purple curve shows a randomly generated time series for an axion mass of $\ma/2\pi = 10^{-3}~\mathrm{Hz}$ ($\tau_c \sim 5~\mathrm{yr}$) for 5 days, including Gaussian white noise with $\sigma_{\rm bkg} = 10^{-2}/\sqrt{\rhodm} \vsun$.  This should be compared to the orange curve, which has the same amplitude and phase in the $\mathbf{e}_z$ direction, but does not include the $\mathbf{e}_x$ and $\mathbf{e}_y$ components as in the 1D Stochastic model of \Eq{eq:1d_stochastic}.  There are clear differences in the time evolution of the axion signal in the 1D and 3D Stochastic cases.  Thus, it is not surprising that attempting to fit a real axion signal using the 1D Stochastic model assumption results in an extremely poor goodness-of-fit as quantified by the reduced chi-squared statistic in the right panel of the figure.

\begin{figure}[t] 
   \centering
   \includegraphics[width=0.5\textwidth]{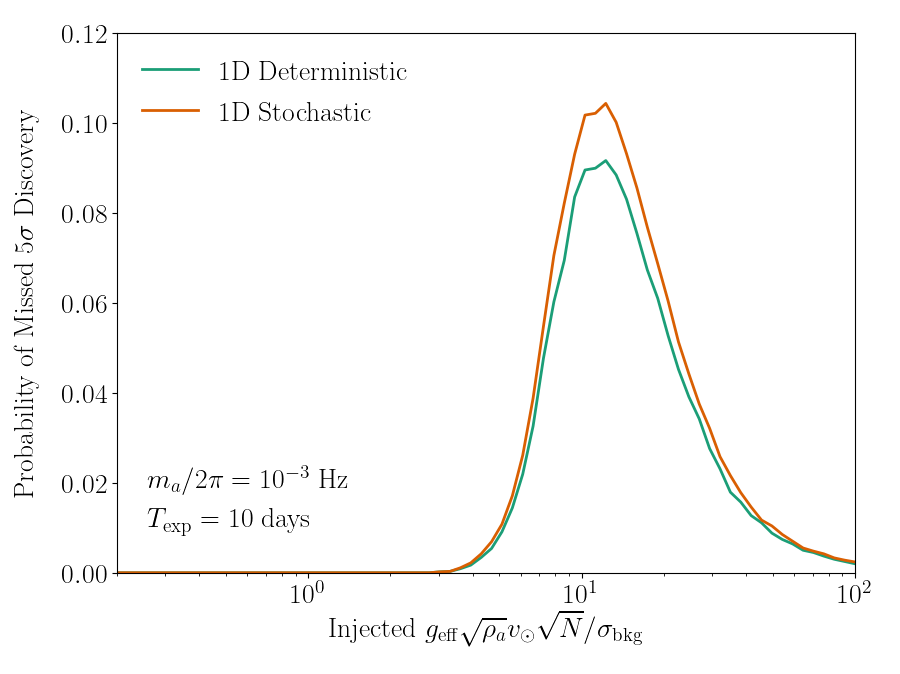}
   \caption{The probability that an axion signal fluctuates such that the full 3D Stochastic analysis yields a local significance $\sqrt{\ts} > 5\sigma$ \textit{and} the 1D Stochastic (\reddish) or 1D Deterministic (\greenish) analyses yield a local significance $\sqrt{\ts} < 3\sigma$. This corresponds to the probability that a $5\sigma$ discovery, which could have been made using the 3D Stochastic analysis, would be ``missed'' in either of the other two benchmark analyses. This probability is plotted as a function of injected signal for the same mock data used in \Fig{fig:signal_injection}.  For injected signals with signal-to-noise ratios of $\sim1$--10, the probability of a missed discovery is $\gtrsim 10\%$ for both the 1D Stochastic and 1D Deterministic analyses. This occurs because the 1D models do not sufficiently capture the time evolution of the axion signal. In particular, the daily modulation of a real axion signal need not follow the form of $m_z(t)$, as shown in \Fig{fig:components}.
   \label{fig:missed_discovery}
   }
\end{figure}

Both the 1D Deterministic and 1D Stochastic models contain no notion of a two-point correlation function. It is therefore not immediately clear how to apply these models to a more general scenario as we considered in \Sec{sec:time_domain_binned}. Although this point alone provides a strong motivation for the use of the full axion model, in the limit of long coherence times, the signal can be treated as coherent and these models can be directly compared to the correct model given by \Eq{eq:coherent}.  

We use the 1D Deterministic and 1D Stochastic models to analyze the same mock data that was used to validate the 3D Stochastic model. For the 1D Stochastic model, the data analysis procedure is exactly analogous to the 3D Stochastic model, except only the $i=z$ parts of \Eq{eq:coherent_fit} and \Eq{eq:coherent_likelihood} are used. For the 1D Deterministic model, the likelihood is different, because the signal amplitude is uniquely determined given the model parameters. In this case, the correct probability distribution for the observed amplitude is a Rice distribution. Changing variables from the observed amplitude and phase to the measured coefficients, $A_z$ and $B_z$, the normalized likelihood is
\be
\mathcal{L}_{\rm 1D,det.}&&(A_z, B_z, \tilde{\sigma}_{z} | g_{\rm eff})  \\
&& = \frac{e^{-(R_z^2 + 2g_{\rm eff}^2\rhodm\vsun^2)/2\tilde{\sigma}^2_{z}}}{2\pi\tilde{\sigma}^2_{z}} \, I_0\left(\frac{R_zg_{\rm eff}\sqrt{2\rhodm\vsun^2}}{\tilde{\sigma}^2_{z}}\right) \,, \nonumber
\ee
where we have introduced the shorthand for the measured amplitude $R_z=\sqrt{A_z^2+B_z^2}$, and $I_0(z)$ is the modified Bessel function of the first kind with order zero.
With this likelihood, the data analysis for the 1D Deterministic model proceeds analogously to the 1D Stochastic and 3D Stochastic model.

We compare the results of the mock data analysis using the three benchmark models in \Fig{fig:signal_injection}. In particular, we find that neither the 1D Stochastic nor the 1D Deterministic models set the correct 95\% upper limit on $g_{\rm eff}$.
Additionally, in the limit of large injected coupling, the 95\% upper limit set by the 1D Deterministic analysis is below the true injected value about 50\% of the time, while the expected frequency of this occurrence is 5\% for a correct model.
More concerning is the different level of local statistical significance as measured by $\ts$ for the three models. The lower panels in \Fig{fig:signal_injection} show that there is considerable more variability in the $\ts$ in the 1D Stochastic and 1D Deterministic models than in the 3D Stochastic model.

We find that there can be considerable difference in $\ts$ across the three models for a given set of mock data.
As shown in \Fig{fig:missed_discovery}, for certain signal-to-noise ratios, both the 1D Deterministic and 1D Stochastic models have a $\sim$10\% chance of finding a test statistic with significance of less than $3\sigma$ \textit{on the same mock data set} where the full analysis finds a significance of at least $5\sigma$.
Furthermore, even when the signal-to-noise ratio is large enough that 1D Deterministic and 1D Stochastic models would claim a significant detection based on the test statistic, these models would fit the data very poorly, as shown in the right panel of \Fig{fig:components}, and such a detection could easily be discarded as spurious.

Thus, using the 1D Stochastic or 1D Deterministic models to analyze a real data set incurs a risk of failing to discover a real axion signal.
This occurs because the 1D models do not fully describe the time evolution of the axion signal. In particular, the daily modulation of a real axion signal need not follow the form of $m_z(t)$, as shown in the left panel of \Fig{fig:components}.
The risk of failing to discover a real axion signal is ameliorated by using the 3D Stochastic model, or the more general model discussed in \Sec{sec:time_domain}.

\section{Conclusions}
\label{sec:conclusions}
Ultralight axion-like particles are well-motivated DM candidates, and their phenomenology in terrestrial direct detection experiments is stochastic.
Any analysis of data from an axion DM direct detection experiment must carefully account for the stochasticity of the axion signal in order to properly interpret the results.
We have shown that, contrary to previous assumptions in the literature, the signal in experiments sensitive to the gradient of the axion field does not always follow the average DM velocity and can be described as a Gaussian random vector field.
We have provided a general framework in which to calculate the two-point correlation functions necessary to fully specify the likelihood for such an axion signal, and we have offered two computationally efficient methods for analyzing a real experimental data set in the time domain, which can be chosen based on the coherence of the axion signal over the lifetime of the experiment.
We illustrate that the analysis procedures that have been used in previous literature can fail dramatically on real data.  Indeed, failing to properly account for the stochastic nature of the axion gradient signal when developing experimental, data acquisition, and data analysis procedures can potentially remove a real signal from the data or else fail to correctly identify a signal from background noise.
As a resource to the interested reader, we have provided an open-source software package at \url{https://github.com/mtmoschella/axionpy} with some features for generating mock axion gradient signals and performing the analysis techniques described in this paper.

\begin{acknowledgments}
  We thank I.~Bloch, K.~Blum, J.~Foster, Y.~Hochberg, L.~Hui, E.~Kuflik, J.~Lee, M.~Romalis, B.~Safdi, and T.~Volansky for useful conversations.  ML is supported by the DOE under Award Number DESC0007968.
  MM is supported by the Simons Foundation.
  WT is supported by the Simons Foundation and the Princeton University Dicke Fellowship.
  The work presented in this paper was performed on computational
resources managed and supported by Princeton Research
Computing, a consortium of groups including the Princeton Institute for Computational Science and Engineering
(PICSciE) and the Office of Information Technology's
High Performance Computing Center and Visualization
Laboratory at Princeton University. This research made use of the \verb|numpy|~\cite{numpy}, \verb|scipy|~\cite{scipy}, \verb|matplotlib|~\cite{matplotlib}, and \verb|astropy|~\cite{astropy1,astropy2} software packages.
\end{acknowledgments}

\appendix
\section*{Appendix}

\setcounter{equation}{0}
\setcounter{figure}{0}
\setcounter{table}{0}
\setcounter{section}{0}
\makeatletter
\renewcommand{\theequation}{A\arabic{equation}}
\renewcommand{\thefigure}{A\arabic{figure}}
\renewcommand{\thetable}{A\arabic{table}}

\section{The Central Limit Theorem}
\label{app:clt}
\noindent\textit{Statement:}\newline
\indent
Let $\{\boldsymbol{x}_n \, | \, n=1,...,N\}$ be a set of $N$ independent and identically distributed random samples from the $k$-dimensional probability density function (PDF) $\mathcal{P}(\boldsymbol{x})$.
Then, in the limit $N\to\infty$, the vector sum
\be
\boldsymbol{S} = \frac{1}{\sqrt{N}}\sum_{n=1}^N \boldsymbol{x}_n \sim \mathcal{N}\left(\sqrt{N}\boldsymbol{\mu},\boldsymbol{\Sigma}\right) 
\label{eq:S}
\ee
is normally distributed with mean $\sqrt{N}\boldsymbol{\mu}$ and covariance matrix $\boldsymbol{\Sigma}$, where
\be
\mu_i &=& \mathrm{E}\left[x_i\right] =  \int \dd^kx\ x_i \, \mathcal{P}(\boldsymbol{x})
\label{eq:clt_mean}
\ee
and
\be
\Sigma_{ij} &=& \mathrm{E}\left[(x_i-\mu_i)(x_j-\mu_j)\right] \el
            &=& \int \dd^kx\ (x_i-\mu_i)(x_j-\mu_j) \, \mathcal{P}(\boldsymbol{x}) \,.
\label{eq:clt_cov}
\ee

\hfill\\

\noindent\textit{Proof:}\newline
\indent
Consider the random variable
\be
\widetilde{\boldsymbol{S}} &\equiv& \boldsymbol{S}-\sqrt{N}\boldsymbol{\mu} \el
&=& \frac{1}{\sqrt{N}}\sum_{n=1}^N \left(\boldsymbol{x}_n-\boldsymbol{\mu}\right) \,,
\label{eq:shifted}
\ee
which is equivalent to $\boldsymbol{S}$, but shifted by a constant.
The characteristic function of $\widetilde{\boldsymbol{S}}$, $\Phi_{\widetilde{\boldsymbol{S}}}(\boldsymbol{t})$, which is defined in terms of the $k$-dimensional Fourier parameter $\boldsymbol{t}$, is\be
\Phi_{\widetilde{\boldsymbol{S}}}(\boldsymbol{t}) &\equiv& \mathrm{E}\left[\exp\left(i\boldsymbol{t}\cdot\widetilde{\boldsymbol{S}}\right)\right] \el
&=& \left(\mathrm{E}\left[\exp\left(\frac{i\boldsymbol{t}\cdot\left(\boldsymbol{x}-\boldsymbol{\mu}\right)}{\sqrt{N}}\right)\right]\right)^N \,,
\ee
where we have used \Eq{eq:shifted} and the fact that the $\boldsymbol{x}_n$ random variables are independent and identically distributed.
Now we expand the exponential in a multivariate Taylor series to leading order in large $N$, i.e., the small parameter $1/\sqrt{N}$, giving
\be
\Phi_{\widetilde{\boldsymbol{S}}}(\boldsymbol{t}) &=& \left(1 -\frac{1}{2}\sum_{i=1}^k\sum_{j=1}^k\frac{\mathrm{E}\left[\left(x_{i}-\mu_i\right)\left(x_{j}-\mu_j\right)\right]t_it_j}{N}\right)^N \el
&=& \left(1- \frac{1}{2N}\boldsymbol{t}^\intercal\boldsymbol{\Sigma}\boldsymbol{t}\right)^N \,.
\label{eq:charfun}
\ee
Note that the first-order term in the Taylor series vanishes due to the fact that $\mu_i=\mathrm{E}[x_i]$.
As $N\to\infty$, \Eq{eq:charfun} becomes
\be
\Phi_{\widetilde{\boldsymbol{S}}}(\boldsymbol{t}) &=& \exp\left[-\frac{1}{2}\boldsymbol{t}^\intercal\boldsymbol{\Sigma}\boldsymbol{t}\right] \,.
\ee
Computing the Fourier inversion, we have the PDF for $\widetilde{\boldsymbol{S}}$,
\be
f_{\widetilde{\boldsymbol{S}}}(\boldsymbol{x}) &=& \frac{1}{(2\pi)^k}\int \dd^k t\ e^{-i\boldsymbol{t}\cdot\boldsymbol{x}}\Phi_{\widetilde{\boldsymbol{S}}}(\boldsymbol{t}) \el
&=& \frac{1}{\sqrt{(2\pi)^k\det\boldsymbol{\Sigma}}}e^{-\frac{1}{2}\boldsymbol{x}^\intercal\left(\boldsymbol{\Sigma}\right)^{-1}\boldsymbol{x}} \,,
\ee
which is the PDF for a multivariate normal distribution with zero mean.  That is,
\be
\widetilde{\boldsymbol{S}}\sim \mathcal{N}\left(0, \boldsymbol{\Sigma}\right) \,.
\ee
Switching back from $\widetilde{\boldsymbol{S}}$ to $\boldsymbol{S}$ by a constant additive transformation, we see that
\be
\boldsymbol{S}\sim \mathcal{N}\left(\sqrt{N}\boldsymbol{\mu},\boldsymbol{\Sigma}\right) \,,
\ee
as required.


\renewcommand{\theequation}{B\arabic{equation}}
\renewcommand{\thefigure}{B\arabic{figure}}
\renewcommand{\thetable}{B\arabic{table}}

\section{Uncertainties from Ordinary Least Squares}
\label{app:ols}
In ordinary least squares (OLS), we fit a $K$-vector of observations $\textbf{y}$ to a linear function of the form
\be
\textbf{y} = \textbf{X}\boldsymbol{\beta} \, ,
\ee
where $\boldsymbol{\beta}$ is an $M$-vector of coefficients, and $\textbf{X}$ is a $K \times M$ matrix.  In particular,  $\textbf{X}$ can be thought of as $M$ different $K$-vectors $\textbf{x}_m$ that are determined by the basis functions that are being fit to. For example, suppose we want to fit a time series
\be
y(t_k) = A\cos(\omega t_k) + B\sin(\omega t_k) \, ,
\ee
where $\omega$ is known and the coefficients $A, B$ are constant with time.  Then, the basis functions are $x_1(t_k) = \cos(\omega t_k)$ and $x_2(t_k) = \sin(\omega t_k)$ and the OLS coefficients are $\beta_1 = A$, $\beta_2 = B$.  The \textit{least squares} solution to this problem is the set of parameters $\boldsymbol{\beta}^*$ that minimize the sum of the squares of the residuals,
\be
\mathrm{SSR}(\boldsymbol{\beta}) = \sum_{k=1}^K \left(y(t_k) - \sum_m \beta_m x_m(t_k) \right)^2 \, .
\ee
The solution $\boldsymbol{\beta}^*$ has the following closed-form expression:
\be
\boldsymbol{\beta}^* = \left(\textbf{X}^\intercal \textbf{X}\right)^{-1}\textbf{X}^\intercal \textbf{y} \,,
\ee
where the $M\times K$ matrix $\textbf{X}^+ \equiv \left(\textbf{X}^\intercal \textbf{X}\right)^{-1}\textbf{X}^\intercal$ is the Moore-Penrose pseudoinverse of $\textbf{X}$.

Our primary goal is to understand the uncertainty on the parameter estimate $\boldsymbol{\beta}^*$, assuming that $\textbf{y}$ has an independent identical Gaussian uncertainty $\delta y$ on each measurement. Assuming $\delta y$ is small enough to propagate errors in the usual way, we see that the uncertainty on the $m^\text{th}$ estimated parameter, $\delta \beta^*_m$, is given by
\be
\left(\delta \beta^*_m\right)^2 = \sum_k \left(X^+_{mk} \, \delta y\right)^2 \, .
\label{eq:general_uncertainty}
\ee

To understand the matrix $\textbf{X}^+$, let us first consider the $M\times M$ matrix $\textbf{X}^\intercal\textbf{X}$, which has elements
\be
(\textbf{X}^\intercal\textbf{X})_{m,m'} = \sum_{k} x_m(t_k)x_{m'}(t_k) \approx K \langle x_m x_{m'} \rangle \, ,
\ee
where we have assumed that the sampling rate is rapid enough and the integration time is long enough to average over the entire domain of the basis functions. If the basis functions $x_m$ satisfy the useful orthogonality property
\be
\langle x_m x_{m'} \rangle = \langle x_m^2\rangle \delta_{m,m'} \,,
\ee
then $\textbf{X}^\intercal\textbf{X}$ is approximately diagonal and the inverse matrix $(\textbf{X}^\intercal\textbf{X})^{-1}$ is trivial to compute. The pseudoinverse matrix is therefore
\be
X^+_{mk} = \sum_{m'} (\textbf{X}^\intercal\textbf{X})^{-1}_{m,m'} x_{m'}(t_k) \approx \frac{x_m(t_k)}{K \langle x_m^2\rangle}
\ee
and so the uncertainties on the least-squares parameters are given by
\be
\left(\delta \beta^*_m\right)^2 \approx \sum_k \frac{x_m^2(t) \, \delta y^2}{K^2\langle x_m^2\rangle^2} \approx \frac{\delta y^2}{K\langle x_m^2\rangle} \,.
\ee
Solving for $\delta \beta^*_m$ explicitly, we obtain
\be
\delta \beta^*_m \approx \frac{\delta y}{\sqrt{K \langle x_m^2\rangle}} \,.
\ee


\renewcommand{\theequation}{C\arabic{equation}}
\renewcommand{\thefigure}{C\arabic{figure}}
\renewcommand{\thetable}{C\arabic{table}}

\section{Evaluating Maxwellian Integrals}
\label{app:max}
This Appendix demonstrates how to compute integrals of the form in \Eq{eq:time_domain_cov_integral} for the specific case of the velocity distribution function in \Eq{eq:maxwell_boltzmann} and under the assumption $\vobsvec(t)\approx \vsunvec$. Changing variables of integration from $\vec{w} \to \vec{v}=\vec{w}+\vsunvec$, we obtain
\be
\langle A_i(t)A_{j}(t') \rangle &=& g_{\rm eff}^2\rhodm \int \dd^3 \vec{v}\ v_iv_j \frac{e^{-(\vec{v}-\vsunvec)^2/2\sigmav^2}}{\left(2\pi\sigmav^2\right)^{3/2}} \cos\left(\Delta \varpi\right) \el
\langle A_i(t)B_{j}(t') \rangle &=& g_{\rm eff}^2\rhodm \int \dd^3\vec{v}\ v_iv_j \frac{e^{-(\vec{v}-\vsunvec)^2/2\sigmav^2}}{\left(2\pi\sigmav^2\right)^{3/2}} \sin\left(\Delta \varpi \right) \el \, 
\ee
where $\Delta\varpi = \frac{1}{2}\ma v^2 (t'-t)$ and the indices $i,j=x,y,z$.

Choosing to evaluate the integral in spherical coordinates $(v,\theta,\varphi)$ with polar axis $\hat{e}_z$, the volume element becomes $\dd^3\vec{v} = v^2\dd v\ \dd\cos\theta\ \dd \varphi$, and the components of $\vec{v}$ are expressed as
\be
v_x &=& v\sin\theta\cos\varphi \el
v_y &=& v\sin\theta\sin\varphi \el
v_z &=& v\cos\theta\, .
\ee
Additionally, $\left(\vec{v}-\vsunvec\right)^2 = v^2 + \vsun^2 - 2v\vsun\cos\theta$ because $\hat{e}_z$ is chosen to be parallel to $\vsunvec$.

Examining the $\dd \varphi$ integrals, it is clear that unless $i=j$, the integral vanishes, and that the $i=j=x$ integrals are equivalent to $i=j=y$. This leaves four unique non-zero integrals to compute:
\be
\langle A_z(t)A_z(t') \rangle &=& g_{\rm eff}^2\rhodm \int_0^\infty \dd v\ v^4 \tilde{f}_{\parallel}(v) \cos\left(\Delta \varpi\right) \el
\langle A_z(t)B_z(t') \rangle &=& g_{\rm eff}^2\rhodm \int_0^\infty \dd v\ v^4 \tilde{f}_{\parallel}(v) \sin\left(\Delta \varpi\right)\el
\langle A_x(t)A_x(t') \rangle &=& g_{\rm eff}^2\rhodm \int_0^\infty \dd v\ v^4 \tilde{f}_{\perp}(v) \cos\left(\Delta \varpi\right) \el
\langle A_x(t)B_x(t') \rangle &=& g_{\rm eff}^2\rhodm \int_0^\infty \dd v\ v^4 \tilde{f}_{\perp}(v) \sin\left(\Delta \varpi\right) \,, \el
\label{eq:four_integrals}
\ee
where we have ignored the escape velocity in order to carry out the $\dd v$ integrals from $0$ to $\infty$.  The $\dd\varphi$ and $\dd\cos\theta$ integrals are encapsulated in the effective speed distributions:
\be
\tilde{f}_{\parallel}(v) &\equiv& \int_{-1}^1 \dd\cos\theta \int_0^{2\pi}\dd\varphi\ \cos^2\theta\frac{e^{-(v^2 + \vsun^2 -2v\vsun\cos\theta)/2\sigmav^2}}{\left(2\pi\sigmav^2\right)^{3/2}} \el
\tilde{f}_{\perp}(v) &\equiv& \int_{-1}^1 \dd\cos\theta \int_0^{2\pi}\dd\varphi\ \sin^2\theta\cos^2\varphi \el
&&\hspace{1.34in} \times \frac{e^{-(v^2 + \vsun^2 -2v\vsun\cos\theta)/2\sigmav^2}}{\left(2\pi\sigmav^2\right)^{3/2}}  \el
\ee
Evaluating these integrals, we obtain the effective speed distributions
\be
\tilde{f}_{\parallel}(v) &\equiv& \frac{\sqrt{2}e^{-(v^2+\vsun^2)/2\sigmav^2}}{\sqrt{\pi}\sigmav^3} \frac{(2+\eta^2)\sinh\eta -2\eta\cosh\eta}{\eta^3}\el
\tilde{f}_{\perp}(v) &\equiv& \frac{\sqrt{2}e^{-(v^2+\vsun^2)/2\sigmav^2}}{\sqrt{\pi}\sigmav^3} \frac{\eta\cosh\eta -\sinh\eta}{\eta^3} \,, \el
\label{eq:eff_speed_dist}
\ee
where $\eta\equiv v\vsun/\sigmav^2$.

Using the results of \Eq{eq:eff_speed_dist} and rewriting the trigonometric functions in \Eq{eq:four_integrals} as complex exponentials, we obtain
\be
\langle A_z(t)A_z(t') \rangle &=& \mathrm{Re}\left\{ \mathcal{Z}_{\parallel} \right\} \el
\langle A_z(t)B_z(t') \rangle &=& \mathrm{Im}\left\{ \mathcal{Z}_{\parallel} \right\} \el
\langle A_x(t)A_x(t') \rangle &=& \mathrm{Re}\left\{ \mathcal{Z}_{\perp} \right\} \el
\langle A_x(t)B_x(t') \rangle &=& \mathrm{Im}\left\{ \mathcal{Z}_{\perp} \right\} \,, \el
\label{eq:complexZ}
\ee
where we have defined the following complex variables
\be
\mathcal{Z}_{\parallel} &\equiv& \sqrt{\frac{2}{\pi}}\frac{g_{\rm eff}^2\rhodm\sigmav^3}{\vsun^3} e^{-\vsun^2/2\sigmav^2} \left(2\mathcal{I}_1 - 2\frac{\vsun}{\sigmav^2}\mathcal{I}_2 + \frac{\vsun^2}{\sigmav^4}\mathcal{I}_3\right) \el
\mathcal{Z}_{\perp}  &\equiv& \sqrt{\frac{2}{\pi}}\frac{g_{\rm eff}^2\rhodm\sigmav^3}{\vsun^3} e^{-\vsun^2/2\sigmav^2} \left(\frac{\vsun}{\sigmav^2}\mathcal{I}_2 - \mathcal{I}_1\right) \,. \el
\ee
The complex Gaussian integrals are defined as:
\be
\mathcal{I}_1 &=& \int_0^\infty \dd v\ v\,\sinh(v\vsun/\sigmav^2) \exp\left[-\frac{v^2}{2\sigmav^2}\zeta\right]\el
\mathcal{I}_2 &=& \int_0^\infty \dd v\ v^2\,\cosh(v\vsun/\sigmav^2) \exp\left[-\frac{v^2}{2\sigmav^2}\zeta\right]\el
\mathcal{I}_3 &=& \int_0^\infty \dd v\ v^3\,\sinh(v\vsun/\sigmav^2) \exp\left[-\frac{v^2}{2\sigmav^2}\zeta\right] \,,\el
\ee
with $\zeta \equiv 1-i\xi$ and $\xi\equiv \ma \sigmav^2 (t'-t)$.  Evaluating the Gaussian integrals, we find that
\be
\mathcal{I}_1 &=& \sqrt{\frac{\pi}{2}} \sigmav\vsun\ \zeta^{-3/2} e^{\vsun^2/2\sigmav^2\zeta} \el
\mathcal{I}_2 &=& \sqrt{\frac{\pi}{2}} \sigmav\vsun^2\ \left(\zeta^{-5/2}+\frac{\sigmav^2}{\vsun^2}\zeta^{-3/2}\right) e^{\vsun^2/2\sigmav^2\zeta} \el
\mathcal{I}_3 &=& \sqrt{\frac{\pi}{2}} \sigmav\vsun^3\ \left(\zeta^{-7/2}+3\frac{\sigmav^2}{\vsun^2}\zeta^{-5/2}\right) e^{\vsun^2/2\sigmav^2\zeta} \el
\ee
and therefore
\be
\mathcal{Z}_{\parallel} &=& g_{\rm eff}^2\rhodm \frac{\zeta\sigmav^2 + \vsun^2}{\zeta^{7/2}}  \exp\left[-\frac{\vsun^2}{2\sigmav^2}\left(1 - \frac{1}{\zeta}\right)\right] \el
\mathcal{Z}_{\perp} &=& g_{\rm eff}^2\rhodm \frac{\sigmav^2}{\zeta^{5/2}} \exp\left[-\frac{\vsun^2}{2\sigmav^2}\left(1 - \frac{1}{\zeta}\right)\right] \,. \el
\ee
Writing the complex variables as a single amplitude and phase, we have
\be
\mathcal{Z}_{\parallel} &=& \mathcal{A}_{\parallel}e^{i\Psi_{\parallel}} \el
\mathcal{Z}_{\perp} &=& \mathcal{A}_{\perp}e^{i\Psi_{\perp}} \,, \el
\ee
where the amplitudes are given by
\be
\mathcal{A}_{\parallel} &=& g_{\rm eff}^2\rhodm  \frac{\sqrt{(\vsun^2 + \sigmav^2)^2 + \sigmav^4\xi^2}}{ \left(1+\xi^2\right)^{7/4}}  \exp\left[-\frac{\vsun^2}{2\sigmav^2}\frac{\xi^2}{1+\xi^2}\right] \el
\mathcal{A}_{\perp} &=& g_{\rm eff}^2\rhodm \frac{\sigmav^2}{ \left(1 + \xi^2\right)^{5/4}} \exp\left[-\frac{\vsun^2}{2\sigmav^2}\frac{\xi^2}{1+\xi^2}\right] \,, \el
\ee
and the phases are given by
\be
\Psi_{\parallel} &=& \frac{\vsun^2}{2\sigmav^2}\frac{\xi}{1+\xi^2} + \frac{7}{2}\arctan\xi - \arctan\left(\frac{\xi\sigmav^2}{\sigmav^2+\vsun^2}\right) \el
\Psi_{\perp} &=& \frac{\vsun^2}{2\sigmav^2}\frac{\xi}{1+\xi^2} + \frac{5}{2}\arctan\xi \, .\el
\ee
Taking the real and imaginary parts according to \Eq{eq:complexZ}, we obtain
\be
\langle A_z(t)A_z(t') \rangle &=& \mathcal{A}_{\parallel}(\xi)\cos\Psi_{\parallel}(\xi) \el
\langle A_z(t)B_z(t') \rangle &=& \mathcal{A}_{\parallel}(\xi)\sin\Psi_{\parallel}(\xi) \el
\langle A_x(t)A_x(t') \rangle &=& \mathcal{A}_{\perp}(\xi)\cos\Psi_{\perp}(\xi) \el
\langle A_x(t)B_x(t') \rangle &=& \mathcal{A}_{\perp}(\xi)\sin\Psi_{\perp}(\xi) \, , \el
\ee
consistent with \Eq{eq:ABcorr}.

\clearpage

\bibliographystyle{apsrev}
\bibliography{comag.bib}
\end{document}